\documentclass{emulateapj}
\usepackage{threeparttable}
\usepackage{natbib}
\usepackage[bookmarks=false, colorlinks=true, citecolor=blue,
  backref=true]{hyperref}              
\hypersetup{
  colorlinks,
  citecolor=blue,
  linkcolor=red,
  urlcolor=blue,
}

\DeclareRobustCommand{\ion}[2]{%
\relax\ifmmode
\ifx\testbx\f@series
{\mathbf{#1\,\mathsc{#2}}}\else
{\mathrm{#1\,\mathsc{#2}}}\fi
\else\textup{#1\,{\mdseries\textsc{#2}}}%
\fi}

\shorttitle{AGC 223218}
\shortauthors{Cao et al.}

\begin{document}
\slugcomment{{\bf Accepted for publication in ApJ}}

\title{Formation of A Nuclear Star Cluster Through A Merger Event In The Low Surface Brightness Galaxy AGC 223218}
\author{Tian-Wen Cao\altaffilmark{1}, Zi-Qi Chen\altaffilmark{1}, Zi-Jian Li\altaffilmark{2,3},
Cheng Cheng\altaffilmark{2,3}, Gaspar Galaz\altaffilmark{4}, 
Venu M. Kalari\altaffilmark{5}, Jun-feng Wang\altaffilmark{1}, Chun-Yi Zhang\altaffilmark{1}, 
Pei-Bin Chen\altaffilmark{1}, Meng-Ting Shen\altaffilmark{1}, Hong Wu\altaffilmark{2}}

\altaffiltext{1}{Department of Astronomy, Xiamen University, 422 Siming South Road, Xiamen 361005, People$'$s Republic of China; \\astrocao@xmu.edu.cn, jfwang@xmu.edu.cn}
\altaffiltext{2}{National Astronomical Observatories, Chinese Academy of Sciences, Beijing 100101, People$'$s Republic of China}
\altaffiltext{3}{Chinese Academy of Sciences South America Center for Astronomy, National Astronomical Observatories, Chinese Academy of Sciences, Beijing 100101, People$'$s Republic of China}
\altaffiltext{4}{Instituto de Astrofisica, Pontificia Universidad  Cat\'olica de Chile,\,$\!$\,Av.$\!$Vicu\~na Mackenna$\!$\,4860,\,7820436 Macul,$\!$\,Santiago, Chile}
\altaffiltext{5}{Gemini Observatory/NSFs NOIRLab, Casilla 603, La Serena, Chile}
\date{27/07/2025}

\begin{abstract}
We present the properties of the nuclear star cluster (NSC) in the low surface brightness galaxy AGC 223218.
The disk of the galaxy can be modeled using two S$\acute{\rm e}$rsic components with distinct central positions:
one representing the inner bright disk and the other corresponding to the extended outer disk.
We estimate the stellar masses of the NSC and the host galaxy using two methods: 
spectral energy distribution (SED) fitting and mass-to-light versus color relations (MLCRs). 
The stellar mass ratio of the NSC to AGC 223218 is 0.094 based on the SED method and 0.072 using MLCRs.
The NSC presents a younger stellar population and a lower [Fe/H] value than the host, 
as determined from SDSS and LAMOST spectra analysis using pPXF fitting.
AGC 223218 is located at the boundary between the Seyfert and star-forming regions in the [SII]-BPT diagram, 
whereas in the [NII]-BPT diagram, it falls in the track of star-forming SDSS galaxies. 
This suggests the presence of strong shocks in AGC 223218.
We propose that the NSC in AGC 223218 may have formed as a result of a merger event.
Furthermore, the observed X-ray luminosity of AGC 223218 with eROSITA
is two orders of magnitude higher than the expected X-ray luminosity from X-ray binaries,
suggesting the presence of an intermediate-mass black hole (IMBH) in the NSC. 
To account for the observed X-ray luminosity, we estimate the IMBH accretion rate to be approximately 0.001.

\end{abstract}


\section{Introduction \label{intro}}
Nuclear Star Clusters (NSCs, \citealt{1983AJ.....88..804C, 1984AJ.....89...64B, 1988ApJ...324..123B,
 2010IAUS..266...58B, 2010ApJ...714..713S, 2011ApJS..196...11S}) 
are among the densest stellar systems in the Universe. NSCs are typically located at the centers of most galaxies, and have stellar masses (M$_*$)
ranging from 10$^4$ to 10$^8$ M$\odot$ and effective radii between 1 and 20 pc \citep{2020A&ARv..28....4N}.
Despite their prevalence, the origin of NSCs remains poorly understood.
Two primary formation scenarios have been proposed: 
the globular cluster (GC) accretion scenario and the \textit{in situ} star formation scenario. 
In the GC accretion scenario, NSCs forms through the gas-free merger of GCs and 
spiral into the galaxy's center due to dynamical friction
\citep{1975ApJ...196..407T, 1976ApJ...203..345T, 1993ApJ...415..616C, 2008ApJ...681.1136C, 
2011ApJ...729...35A, 2014ApJ...785...71G, 2014MNRAS.444.3738A, 2022MNRAS.511.1860S}.
This mechanism can explain the metal-poor NSCs observed in some dwarf galaxies
 \citep{2020A&A...634A..53F, 2020MNRAS.495.2247J, 2022A&A...667A.101F}. 
In the \textit{in situ} star formation scenario, NSCs form directly from gas at the galaxy's center 
 \citep{1982A&A...105..342L, 2004ApJ...605L..13M, 2006ApJ...642L.133B,
 2007A&A...462L..27S, 2015ApJ...812...72A}. 
 This scenario suggests that host galaxies have complex star formation histories, 
 and the resulting NSCs tend to exhibit younger ages. 

Intermediate-mass black holes (IMBHs, \citealt{2017IJMPD..2630021M, 2019NatAs...3..755W}) 
can reside within NSCs \citep{1975ApJ...200L.131S, 2013ARA&A..51..511K, 2014ApJ...796...40M, 2023MNRAS.520.5964Y}.
With the increasing availability of X-ray data, such as from the 
extended ROentgen Survey with an Imaging Telescope Array All-Sky Survey
(eROSITA, \citealt{2012arXiv1209.3114M, 2021A&A...647A...1P, 
2021A&A...656A.132S}),
numerous active galactic nuclei (AGN) in dwarf galaxies have been identified
(e.g. \citealt{2024MNRAS.527.1962B, 2024ApJ...974...14S}).
Dwarf galaxies offer crucial insights into black hole (BH) seeding mechanisms due to their relatively 
simple merger and accretion histories. The BHs within these galaxies represent some of 
the closest analogs to primordial seed black holes 
\citep{2022NatAs...6...26R, 2020ARA&A..58..257G, 2023MNRAS.518.1880B}.
Additionally, NSCs may serve as valuable proxies for studying central 
black holes and their co-evolution with their host galaxies.

Detailed studies of NSCs have predominantly 
focused on high surface brightness galaxies (HSBGs). 
Researches on NSCs in low surface brightness galaxies (LSBGs, 
\citealt{1987AJ.....94...23B, 2015ApJ...798L..45V, 2024Univ...10..432C}) 
remain scarce, with only a few works addressing this topic (e.g. 
 \citealt{2023AAS...24140224L, 2024AJ....167...61L, 2024AJ....168...45K}).
LSBGs hosting NSCs tend to lie on 
 the galaxy red sequence and show a higher occurrence in denser environments \citep{2024AJ....167...61L}. 
 However, distance uncertainties of candidate samples and the lack of 
 spectroscopic observations of NSCs in LSBGs significantly limit current research.
LSBGs have a shortage of molecular gas \citep{2017AJ....154..116C, 2024ApJ...971..181C, 2024ApJ...975L..26G},
low star formation rate \citep{2019ApJS..242...11L},
low surface stellar mass densities \citep{1996ApJ...469L..89D}, low metallicity \citep{2023ApJ...948...96C},
and significant dark matter domination \citep{1997ASPC..117...39D}. 
These properties create an environment significantly different from that of HSBGs, 
offering a unique opportunity to investigate the mechanisms driving NSC formation under varying conditions, 
the influence of dark matter halos on NSC properties, and the star formation history (SFH) of LSBGs.

In this paper, we present the properties of the NSC hosted by a LSBG AGC 223218.
The observational data are described in Section 2, the analysis and results are presented in Section 3, 
and the discussion is provided in Section 4.
We adopt a distance of 18.8$^{-3.78}_{+3.09}$ Mpc for AGC 223218 \citep{2018ApJ...861...49H}, 
where 1$''$ corresponds to 91 pc, ensuring consistency across the estimations in this paper.
The superscript and subscript values represent the difference between this distance and the minimum/maximum 
values provided by the NASA/IPAC Extragalactic Database (NED).

\section{Observational data of AGC 223218}
\begin{figure}
  \begin{center}
  \includegraphics[width=3.3in]{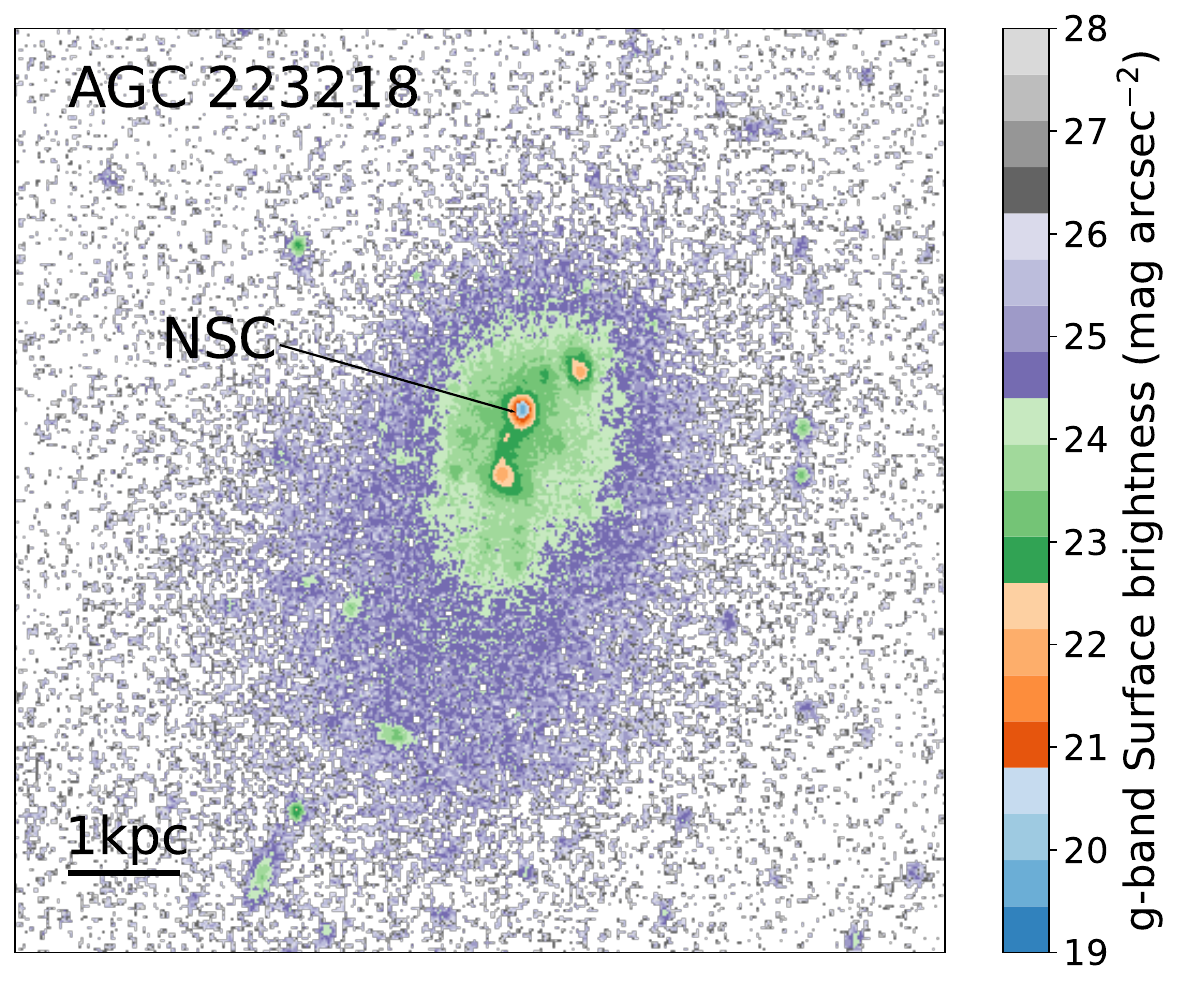}
  \end{center} 
\caption{DECaLS g band surface brightness of AGC 223218. 
The galaxy exhibits the asymmetric structure. 
Three main clumps (including the NSC, the brightest one) 
in the brighter disk, and other clumps are further out.}
   \label{surface}
\end{figure}

We have selected a galaxy sample by cross-matching the eROSITA/eRASS1 main catalog \citep{2024A&A...682A..34M} 
with the Arecibo Legacy Fast ALFA (ALFALFA, \citealt{2005AJ....130.2598G}) Extragalactic HI Source Catalog \citep{2018ApJ...861...49H} 
to investigate different gas components' properties in galaxies. 
Among the selected galaxies, we identified AGC 223218, a LSBG featuring a bright NSC as shown in Figure\,\ref{surface}.
We performed a multi-band analysis of this system using the following publicly available data.

\subsection{Imaging data}

The ultraviolet (UV) data is from the Galaxy Evolution Explorer (GALEX, \citealt{2005ApJ...619L...1M}), 
including FUV and NUV bands. The angular resolutions of FUV and NUV are 4.2$''$ and 5.3$''$, respectively.

The optical data are from the Sloan Digital Sky Survey data release 12 (SDSS DR12, \citealt{2015ApJS..219...12A}),
which includes the u, g, r, i, and z bands, and from the Dark Energy Camera Legacy Survey (DECaLS, \citealt{2019AJ....157..168D}) 
including g, r, and z bands.
The median point spread functions (PSFs) of the r-band are approximately 1.4$''$ for SDSS and 1.2$''$ for DECaLS.

The near-infrared data are from the UKIRT Infrared Deep Sky Survey 
(UKIDSS, \citealt{2007MNRAS.379.1599L}), covering the Z, Y, J, H, and K bands with a typical PSF of $\sim$1.0$''$,
 and from the Wide-field Infrared Survey Explorer 
(WISE, \citealt{2010AJ....140.1868W}), which includes W1, W2, W3, and W4 bands. 
The angular resolutions of W1 and W2 are 6.1$''$ and 6.4$''$, respectively.

\subsection{X-ray data}

The X-ray data are obtained from eROSITA. 
The on-axis PSF of eROSITA at 1.5 keV is about 15$''$.
X-ray data reveal a point source associated with the system
 in the 0.2-2.3 keV band \citep{2024A&A...682A..34M}.

\begin{figure*}
  \begin{center}
  \includegraphics[width=6.5in]{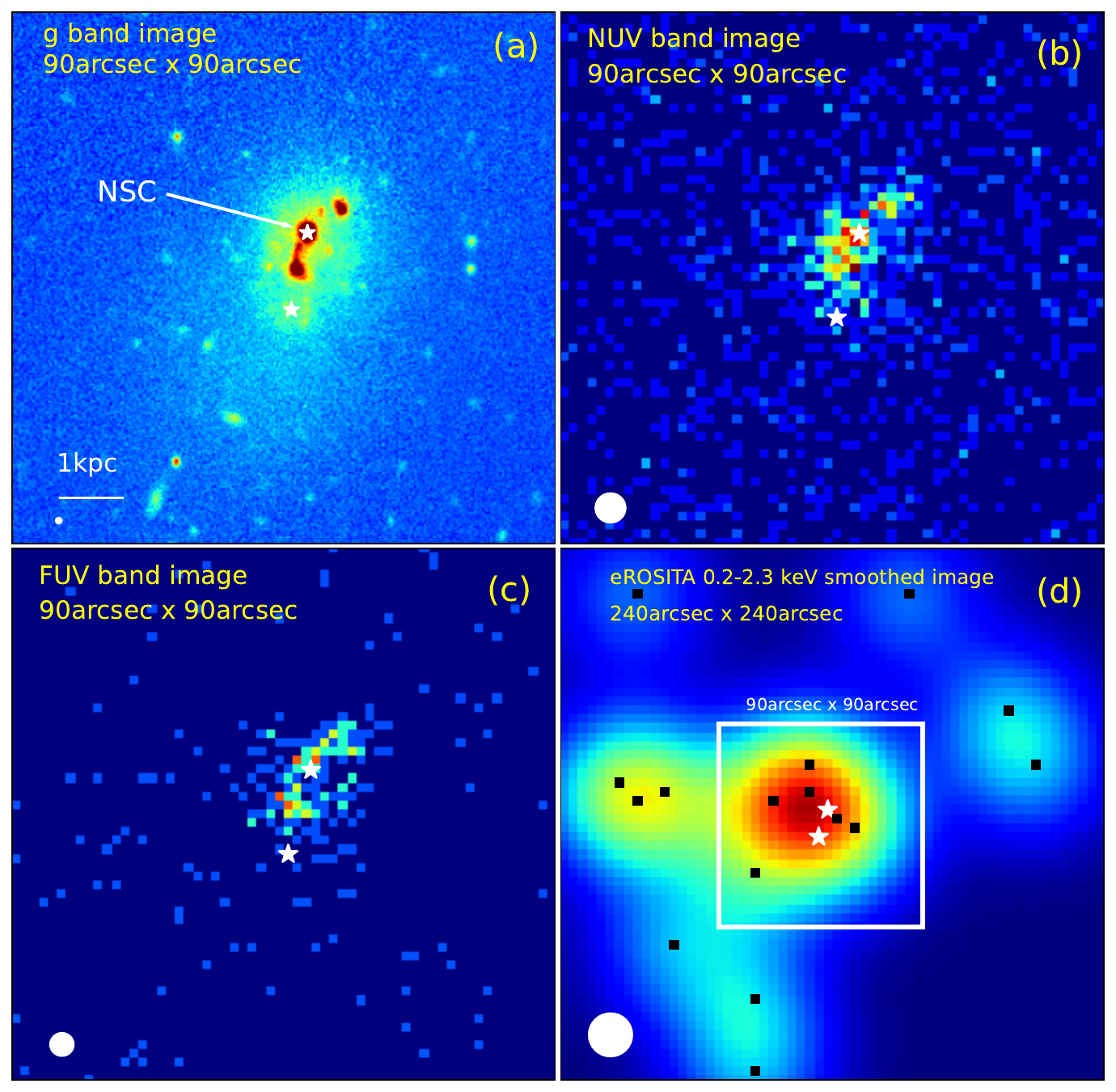}
  \end{center}
\caption{Multi-band images of AGC 223218. 
Panel (a): g band image from DECaLS DR9. 
Panel (b): NUV band image. Panel (c): FUV band image. Panel (d): eROSITA 0.2-2.3 keV image.
The white-filled circle in the left corner of each panels represents the resolution. 
The field of view of g band, NUV, and FUV is 90$''$ x 90 $''$.
For X-ray data, the color code image is the smoothed image with a Gaussian kernel profile with $\sigma$ = 24$''$, and
the X-ray counts are shown with black (one count) pixels. 
The white square in the X-ray image is the same region (90$''$ x 90 $''$) of the g band image.
Two white stars represent the fitting centers of the double-S$\acute{\rm e}$rsic components of the g band image,
which are the same as Figure\,\ref{galfit}.}  
    \label{image}        
\end{figure*}

The multi-band images are presented in Figure\,\ref{image}. 
The morphologies in the FUV, NUV, and g band images are consistent within the brighter disk around the NSC 
and primarily exhibit clumpy structures.

\subsection{Spectroscopy data}
In this system, the NSC has a 3$''$ fiber spectrum from SDSS. 
The host galaxy, AGC 223218, has a 3.3$''$ fiber spectrum obtained
from the Large Sky Area Multi-Object Fiber Spectroscopic Telescope (LAMOST; 
also called the Guo Shou Jing Telescope; \citealt{2012RAA....12.1197C, 2015RAA....15.1095L}) 
and an HI spectrum from the ALFALFA.
The HI mass (M$_{\rm HI}$) of AGC 223218 is (8.7$\pm$2.1)$\times$10$^7$ M$_{\odot}$, with 
50\% HI line peak width (W$_{50}$) of 52 km s$^{-1}$ \citep{2018ApJ...861...49H}.

\section{Data analysis and results}

\subsection{Optical Morphology and system velocity}
\begin{figure*}
  \begin{center}
  \includegraphics[width=7.0in]{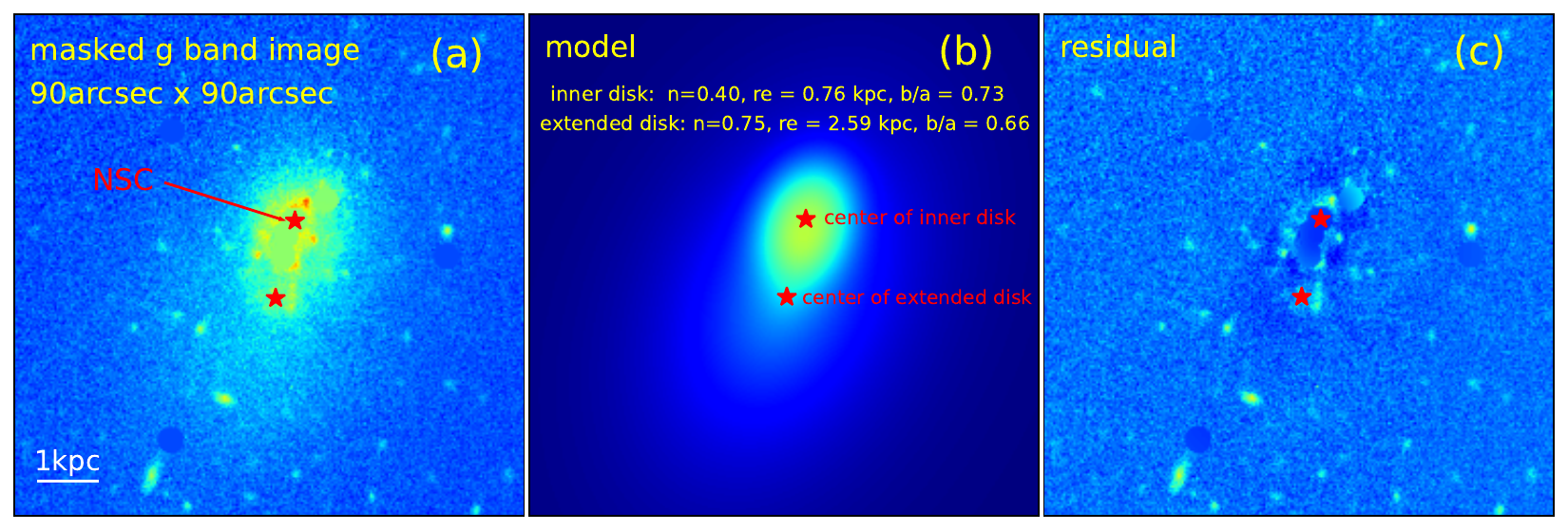}
  \end{center} 
\caption{Panel (a): the final masked g band image from DECaLS. Panel (b): 
the double-S$\acute{\rm e}$rsic model image. Panel (c): the residual image. 
Two red stars represent the fitting centers of the double-S$\acute{\rm e}$rsic components.}
   \label{galfit}
\end{figure*}

We fit the disk of AGC 223218 in the DECaLS image using 
GALFIT \citep{2002AJ....124..266P} with a free double-Sersic model combined with a sky component.
Since the bright clumpy structures significantly affected the disk fitting,
we identified point-like structures using the DAOFIND algorithm \citep{1987PASP...99..191S} 
and masked them with a circular shape. 
The masked regions were filled with values equal to 3$\sigma$, 
where $\sigma$ was the standard deviation of the nearby 25$''$ $\times$ 25$''$ box region.
The final masked g band image is shown in Figure\,\ref{galfit}(a).

The disk of AGC 223218 is well fitted with double-S$\acute{\rm e}$rsic components: 
an inner disk surrounding the NSC and an extended outer disk. 
The fitting results are presented in Figure\,\ref{galfit}.

The double-S$\acute{\rm e}$rsic components have distinct centers. 
The center of the inner disk coincides with the position of the NSC.
The inner disk exhibits a S$\acute{\rm e}$rsic index of n = 0.40 
with an effective radius (R$\rm_e$) of 0.76 kpc, 
while the extended disk shows n = 0.75 with R$\rm_e$ = 2.59 kpc.
Both components exhibit a similar axis ratio of approximately 0.7.
The surface brightness near the inner disk is higher than the extended disk, as shown in Figure\,\ref{surface}. 
The FUV and NUV images reveal similar structures, indicating active star formation in the inner disk.

\begin{figure}
  \begin{center}
  \includegraphics[width=3.5in]{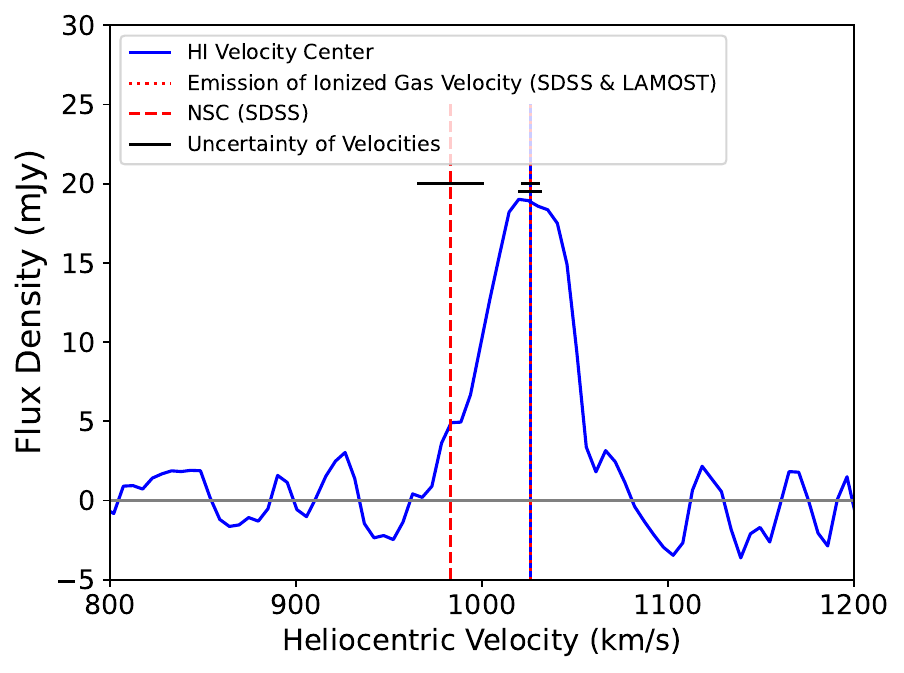}
  \end{center} 
\caption{The HI spectrum from ALFALFA. The blue solid and red dotted lines represent 
the heliocentric velocity center of the HI gas and emissions of ionized gas.
The red dashed line represents the NSC heliocentric velocity from NED.
The black lines are velocity uncertainties of the velocities. 
   \label{HI}}
\end{figure}

We show the HI spectrum in Figure\,\ref{HI}.
The heliocentric velocity of the NSC is 983$\pm$18 km s$^{-1}$, derived from SDSS,
while the HI gas center has a heliocentric velocity of 1026$\pm$5 km s$^{-1}$ \citep{2009A&A...506..677H}.
The spectrum of the NSC in AGC 223218 displays weak [OIII] and H$\alpha$ emissions.
We compared the emission velocities of both [OIII] and H$\alpha$ from the SDSS and LAMOST fiber spectra 
and found that they are consistent with the heliocentric velocity of the HI gas as shown in Figure\,\ref{HI}. 
This indicates a velocity offset of -43 km s$^{-1}$ for the NSC relative to the gas components.
This offset is possibly caused by the uncertain measurement 
from the low signal-to-noise ratio of the SDSS fiber spectrum.
Additionally, the offset between the stellar and gas components can result from
galaxy mergers or tidal disturbances \citep{2015A&A...582A..21B, 2016A&A...595A..56S, 2022NatAs...6.1464C}, 
star formation feedback\citep{2019NatAs...3...82C}, or other energetic processes within galaxies.

\subsection{Stellar mass}
We employed two methods, spectral energy distribution (SED) fitting 
and mass-to-light versus colour relations (MLCRs), to estimate the stellar masses (M$_{*}$) 
of AGC 223218 and its NSC.

The stellar masses and star formation rate (SFR) were derived using SED method 
by CIGALE software \citep{2019A&A...622A.103B}, 
based on the \cite{2003MNRAS.344.1000B} (hereafter BC03) stellar population models 
and the \cite{2003PASP..115..763C} initial mass function (IMF). For AGC 223218,
nine photometric bands covering 0.15-4.6 $\mu$m were used: 
GALEX (FUV, NUV), SDSS (u, g, r, i, z), and WISE (W1, W2). For the NSC,
twelve photometric bands over a similar wavelength range were applied,
incorporating GALEX (FUV, NUV), SDSS (u, g, r, i, z), and UKIDSS (Z, Y, J, H, K).
As the NSC is unresolved in the photometric images, we adopted a 4.2$''$ and 5.3$''$ 
aperture for FUV and NUV, respectively, and a 3$''$ aperture for the remaining bands.
The stellar masses of AGC 223218 (M$^{\rm SED}_{\rm *gal}$) is estimated to be (5.2$\pm$1.9)$\times$10$^7$ M$_{\odot}$,
with a corresponding SFR of 0.013 $\pm$ 0.005 M$_{\odot}$ yr$^{-1}$. 
The stellar masses of the NSC (M$^{\rm SED}_{\rm *NSC}$) is (4.9$\pm$1.6)$\times$10$^6$ M$_{\odot}$,
yielding a mass ratio M$^{\rm SED}_{\rm *NSC}$/M$^{\rm SED}_{\rm *gal}$ of 0.094.

The MLCR formula used is adopted from \cite{2015MNRAS.452.3209R}, expressed as:
\begin{equation}
  \rm log(M_*/L_*)_g = 2.029 \times (g - r) - 0.984
\end{equation}  
based on the BC03 stellar population models and the \cite{2003PASP..115..763C} IMF. 
We used DECaLS g and r bands to derive the color.
The flux of AGC 223218 was measured within the effective radius of the extended disk, 
while the flux of the NSC was measured within a 3$''$ aperture (PSF$_g$ = 1.35$''$).
The (g-r) color is 0.461$\pm$0.010 mag for the NSC and 0.478$\pm$0.014 mag for AGC 223218.
The stellar mass of AGC 223218 (M$^{\rm MLCR}_{\rm *gal}$) is (1.2$\pm$0.08)$\times$10$^8$ M$_{\odot}$,
and the stellar mass of the NSC (M$^{\rm MLCR}_{\rm *NSC}$) is (8.7$\pm$0.4)$\times$10$^6$ M$_{\odot}$
This yields a mass ratio M$^{\rm MLCR}_{\rm *NSC}$/M$^{\rm MLCR}_{\rm *gal}$ of 0.072.

\begin{figure}
  \begin{center}
  \includegraphics[width=3.3in]{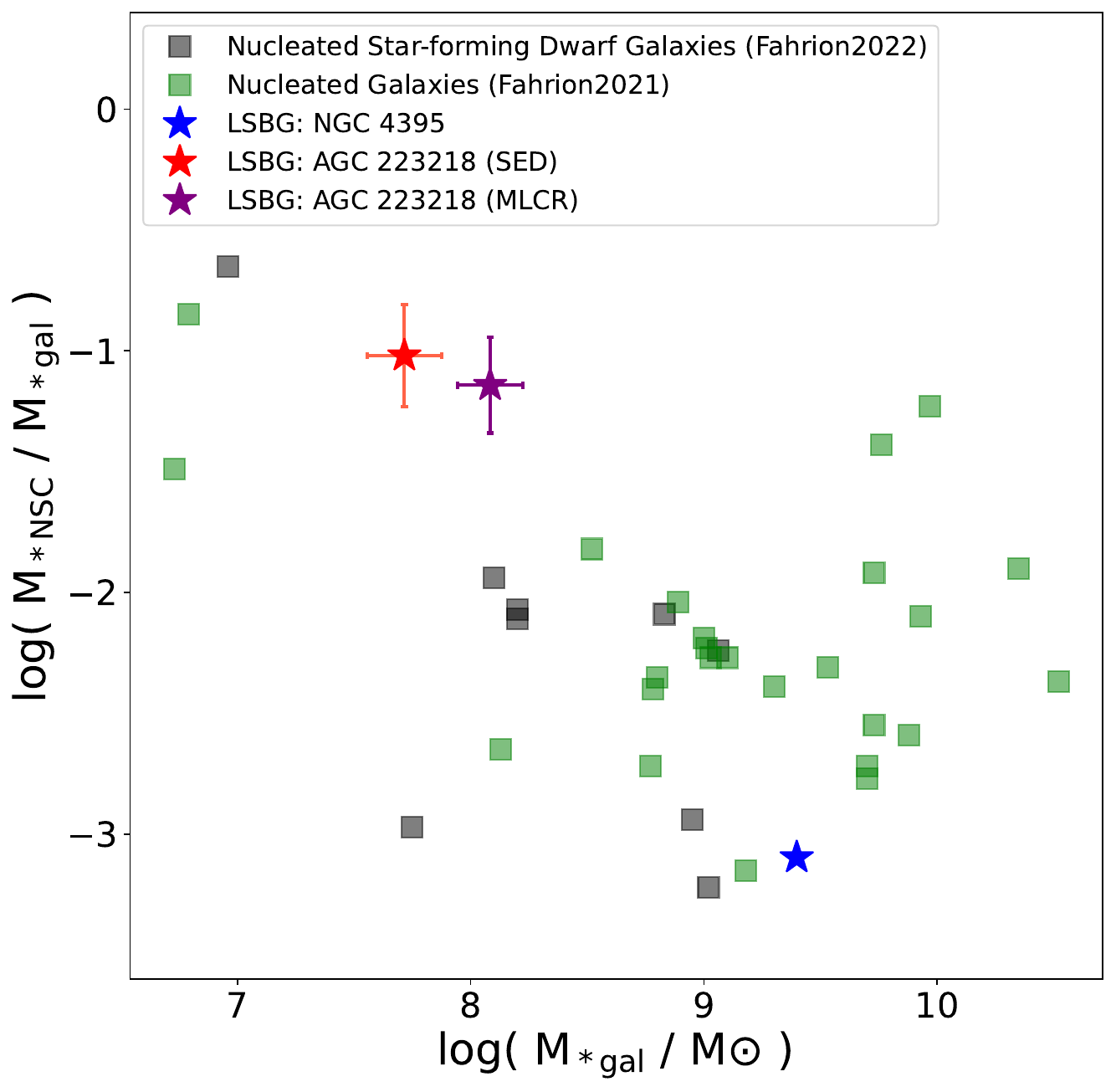}
  \end{center} 
\caption{Diagram of the stellar mass of host galaxy (M$_{\rm *gal}$) vs. the mass ratio of NSC to the host
M$_{\rm *NSC}$/M$_{\rm *gal}$. The green and gray squares are nucleated galaxies and 
nucleated star-forming dwarf galaxies from \cite{2021A&A...650A.137F} and \cite{2022A&A...667A.101F}.
The red and purple stars represent AGC 223218, with stellar masses 
derived using the SED fitting and MLCR methods, respectively. 
The blue star denotes NGC 4395, another LSBG.}
The stellar mass of NGC 4395 and its NSC is from \cite{2020AstBu..75..234S} 
and \cite{2015ApJ...809..101D}, respectively.
   \label{ratio}
\end{figure}

We compare the mass ratio between NSCs and their host galaxies in Figure\,\ref{ratio}.
Nucleated galaxies exhibit a wide range of M$_{\rm *NSC}$/M$_{\rm *gal}$ ratio, with
most NSCs contributing less than one percent of the total galaxy mass. 
The NSC in AGC 223218 exhibits a higher mass fraction in Figure\,\ref{ratio}.
The M$_{\rm *NSC}$/M$_{\rm *gal}$ ratio shows no clear correlation with M$_{\rm *gal}$.
The variation M$_{\rm *NSC}$/M$_{\rm *gal}$ ratio may be influenced by different 
formation mechanisms and evolutionary histories.
The high mass fraction of the NSC in AGC 223218 indicates its special formation mechanism.

\subsection{Stellar population}

The spectrum of NSC shows characteristic spectral lines of A-type stars.
Taking advantage of the SDSS fiber spectrum of the NSC,
we use Penalized Pixel-Fitting (pPXF, \citealt{2004PASP..116..138C}) 
method to fit the spectrum in a wavelength range between 4000$\rm \AA$ and 9000$\rm \AA$.
The stellar population models used the BC03 and adopted the \cite{2003PASP..115..763C} IMF.
The light-weighted average stellar age is 724 Myr, and
the light-weighted mean metallicity [Fe/H] is 0.12 for the NSC.

\begin{figure*}
  \begin{center}
  \includegraphics[width=7.0in]{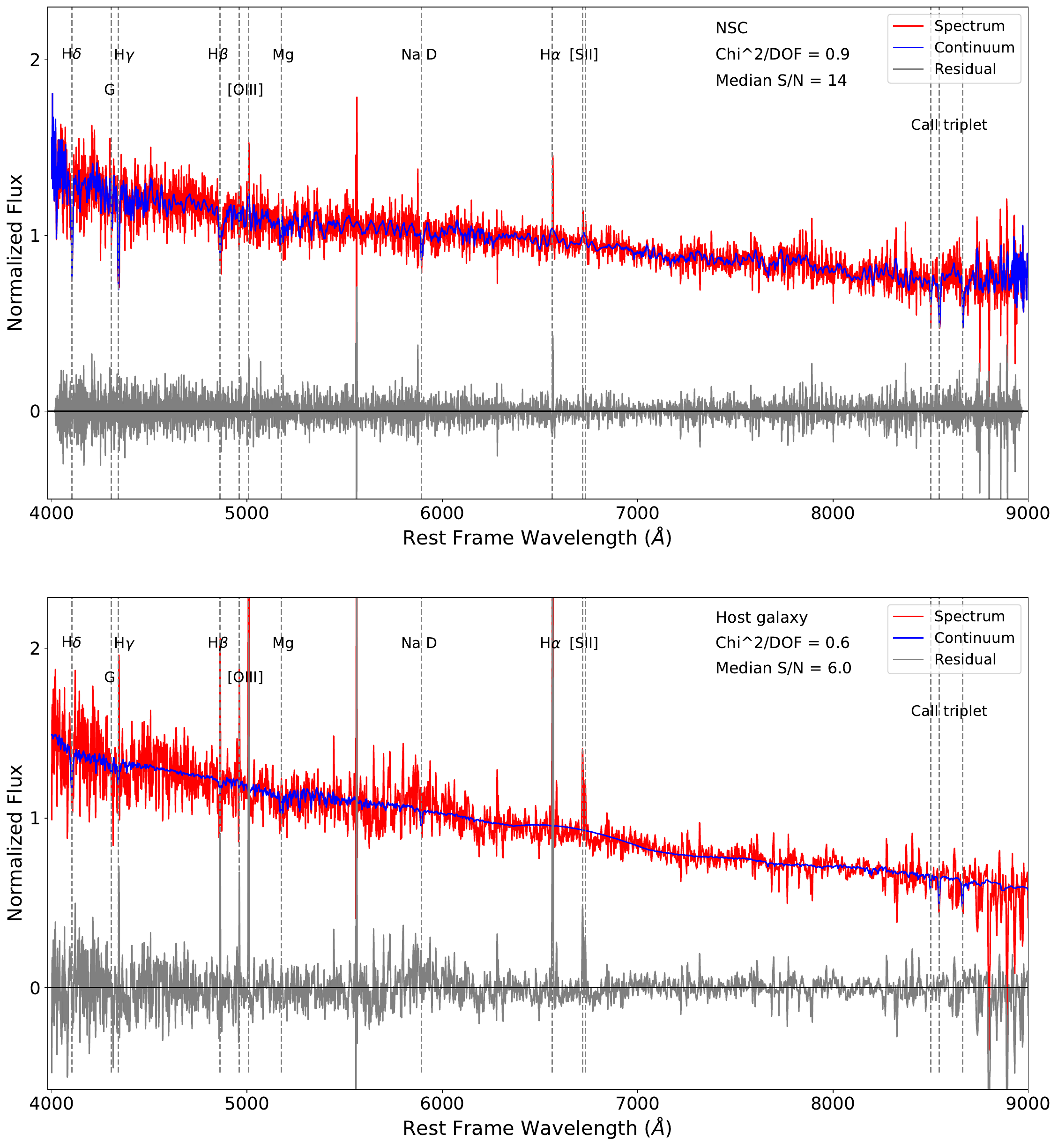}
  \end{center}
\caption{Upper panel: The SDSS fiber spectrum of the NSC position in an aperture of a radius of 1.5$''$
(shown in red) and with the continuum from pPXF fitting (shown in blue). The gray line represents the residual.
The dashed gray lines mark different lines$'$ positions. Lower panel: 
The smoothed LAMOST fiber spectrum of the host galaxy in an aperture of a radius of 1.65$''$ (shown in red) 
and with the continuum from pPXF fitting (shown in blue). The gray line represents the residual.
The dashed gray lines mark different lines$'$ positions.}
    \label{spec}        
\end{figure*}

We used the same fitting parameters for the LAMOST spectrum to derive the properties of the host galaxy.
A one-dimensional boxcar smoothing filter with a width of 4 pixels was applied 
to enhance the signal-to-noise ratio (S/N) of the spectrum. 
The light-weighted average stellar age of the host is 1.73 Gyr, and
the light-weighted mean metallicity [Fe/H] is 0.26.
The modeled spectra from pPXF fitting of the NSC and AGC 223218 are shown in Figure \ref{spec},
 and the age-weighted fractional contributions of stellar populations are listed in Table \ref{table:age}.

We categorize the stellar population contributions into two distinct age groups: 
 a young stellar population (age $<$ 1 Gyr) and an old stellar population (age $>$ 1 Gyr). 
The results in Table \ref{table:age} show that the young stellar population accounts for
69.7\% of the nuclear star cluster (NSC) and 39.9\% of the total stellar population in AGC 223218.
The NSC in AGC 223218 presents a younger stellar population and lower metallicity
than the host.

\begin{table}[h!tb]
  \caption{The age-weighted fractional contributions of stellar populations} 
  \centering 
  \begin{tabular}{c c c} 
  \hline\hline
   1  & 2 & 3\\ 
  \hline
  Object & Age &  Fraction \\
         & (Gyr) &  (\%)    \\
 \hline
        &  0.01 &  10.4 \\ 
  NSC   &  0.50 &  59.3 \\
         &  6.31 &  30.3 \\
\hline
           & 0.01  & 1.5 \\
            & 0.03  & 23.2\\
AGC 223218 & 0.25  & 15.2\\
            & 6.31  & 2.7 \\
            & 15.85 & 57.4\\ 
\hline
  \end{tabular}
  \label{table:age} 
\end{table}

\begin{figure*}
  \begin{center}
  \includegraphics[width=6.5in]{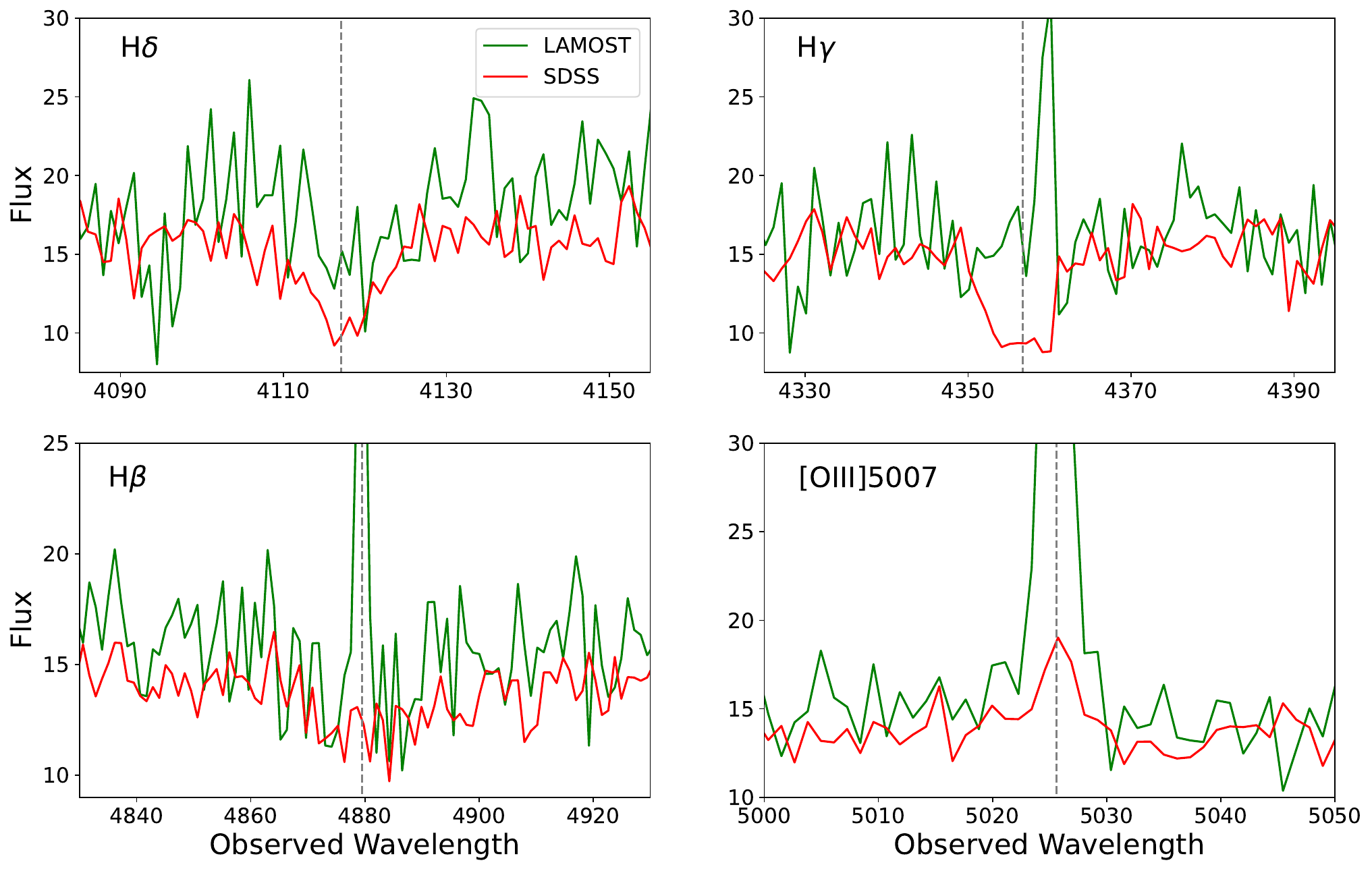}
  \end{center}
\caption{Comparison of H$\delta$, H$\gamma$, H$\beta$, and [OIII]$\lambda$5007 features
between the spectra from LAMOST (green line) and SDSS (red line).
The grey dashed line is the heliocentric velocity center of HI gas.}  
    \label{lines}        
\end{figure*}

We note that the LAMOST data provide a fiber positioned to observe the NSC, 
with a pointing precision of 0.4$''$ \citep{2018ApJS..238...30Z}. 
The pointing offset between the LAMOST and SDSS fiber positions is 1$''$.
The LAMOST spectrum exhibits characteristics of a star-forming galaxy in the BPT diagram,
showing stronger emission lines (e.g., [OIII], H$\beta$, H$\alpha$) and weaker absorption 
lines (e.g., H$\delta$, H$\gamma$) compared to the SDSS spectrum. 
This suggests that the LAMOST spectrum likely originates from a nearby star-forming region rather than the NSC itself, 
possibly due to the pointing offset.
We present a comparison between the SDSS and LAMOST spectra in Figure\,\ref{lines}. 
The NSC spectrum is not expected to exhibit strong emissions. 
\cite{2022A&A...667A.101F} present the spectroscopic analysis of 
a sample of nine nucleated dwarf galaxies (10$^7$ $<$ M$_{*gal}$ $<$ 10$^9$ M$_\odot$)
and all NSC spectra show emission lines.
\cite{2022A&A...667A.101F} explains that emissions are not always linked to star formation 
within the NSC but may originate from other regions along the line of sight. 

\subsection{BPT diagram}
\begin{figure}
  \begin{center}
  \includegraphics[width=3.3in]{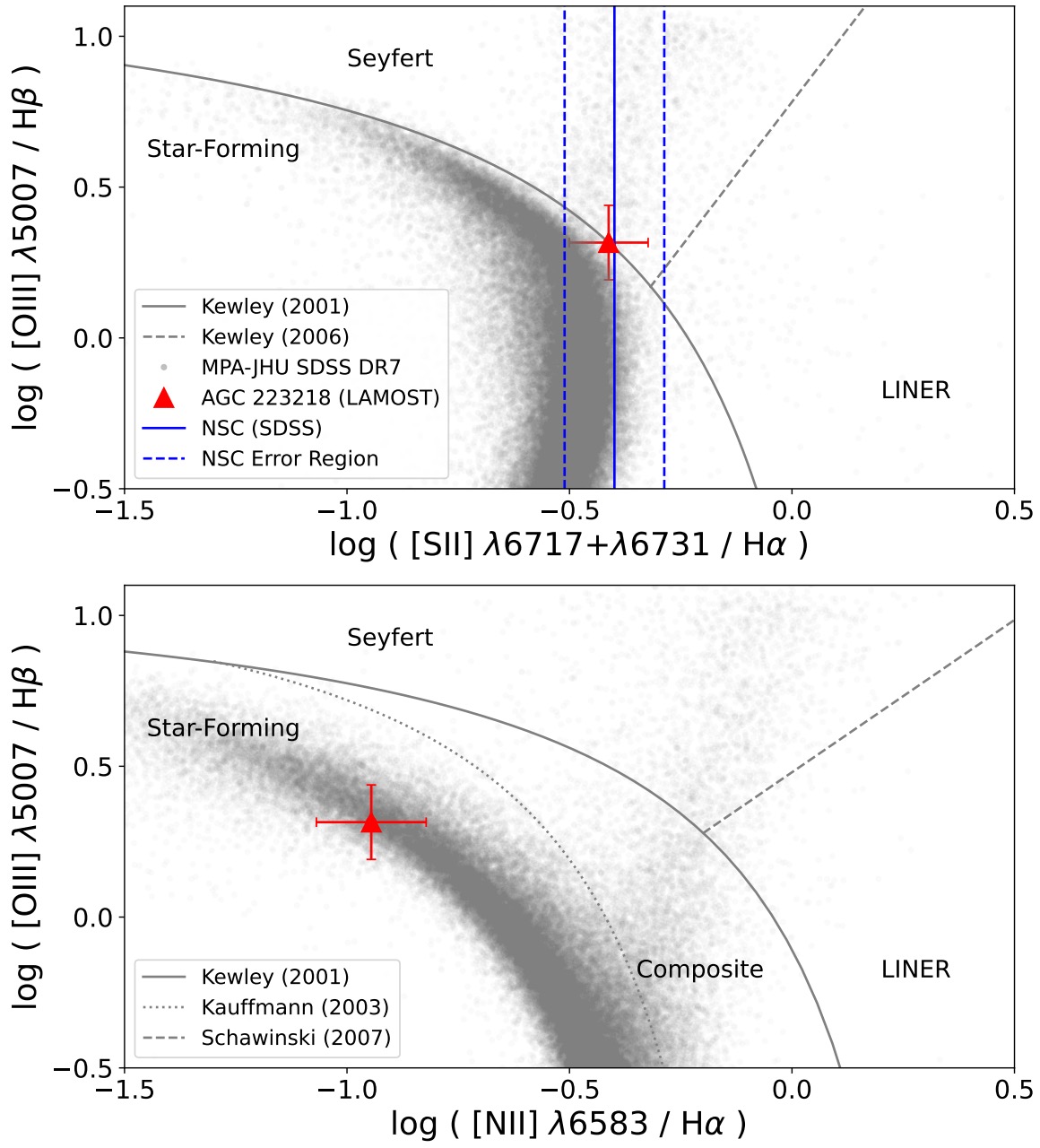}
  \end{center} 
\caption{Upper Panel: The [SII]-BPT diagram. The gray solid line and the gray dashed line are from \cite{2001ApJ...556..121K} and 
\cite{2006MNRAS.372..961K}, respectively. Gray dots represent galaxies from the MPA-JHU SDSS DR7 \citep{2004MNRAS.351.1151B}.
The red triangle represents AGC 223218 according to the LAMOST spectrum. 
The blue line represents the X-axis value of NSC according to the SDSS spectrum and the blue dashed lines are 
the error of the X-axis value. We note that the emissions in the NSC spectrum are possible from the host galaxies.
Lower Panel: The [NII]-BPT diagram. The gray solid line is from \cite{2001ApJ...556..121K}, 
the dotted line is from \cite{2003MNRAS.346.1055K}, and the dashed line is from \cite{2007MNRAS.382.1415S}.
\label{BPT}}
\end{figure}

The Baldwin-Phillips-Terlevich (BPT) diagram is a powerful tool 
for characterizing extragalactic properties based on
various emission-line strength ratios \citep{1981PASP...93....5B}.
We measured the emission lines H$\beta$, [OIII]$\lambda$5007, H$\alpha$, [NII]$\lambda$6583, 
[SII]$\lambda$6717, and [SII]$\lambda$6731 by fitting the LAMOST spectrum with a model that includes the stellar continuum, 
absorption features, and gas emission lines simultaneously.
Figure\,\ref{BPT} presents the BPT classification of AGC 223218.
In the [SII]-BPT diagram, AGC 223218 is located at the boundary between 
the Seyfert and star-forming regions,
whereas in the [NII]-BPT diagram, it falls in the star-forming track of SDSS galaxies. 
Since the [SII] doublet is enhanced in shocked gas, this suggests that strong shocks are responsible 
for the deviation of AGC 223218 from the star-forming track observed in SDSS.
The SDSS spectrum of the NSC does not exhibit H$\beta$ emission and only weak emission lines. 
Therefore, we determine the x-axis position of the NSC in the [SII]-BPT diagram. 
Notably, the ([SII]/H$\alpha$) ratio measured from the SDSS spectrum is consistent
with that obtained from the LAMOST spectrum.
The emissions in the NSC spectrum are possible from the host galaxies.

\subsection{X-ray Luminosity}
The X-ray image in 0.2-2.3 keV shows a weak detection, as illustrated in Figure\,\ref{image}.
The eROSITA/eRASS1 main catalog classifies AGC 223218 as a point source (PS), 
with the X-ray detection coinciding spatially with the NSC \citep{2024A&A...682A..34M}.
The X-ray luminosity in 0.2-2.3 keV (L$_{\rm 0.2-2.3keV}$) is (2.44$\pm$1.19)$\times$10$^{39}$ erg s$^{-1}$,
as derived from the eROSITA/eRASS1 main catalog, assuming a spectral power-law slope of 2.0 
and an absorbing column density of N$_{\rm H}$ = 3$\times$10$^{20}$ cm$^{-2}$, a typical value of the Galactic absorption \citep{2024A&A...682A..34M}.
To estimate the X-ray luminosity in the 2-10 keV band (L$_{2.0-10keV}$),
we applied the band ratio conversion from \cite{2025A&A...694A.128K}, 
resulting in L$_{\rm 2.0-10keV}^{\rm obs}$ = (2.55$\pm$1.24)$\times$10$^{39}$ erg s$^{-1}$.

X-ray binaries (XRBs) are stellar systems consisting of a compact object—such as a neutron star, 
stellar-mass black hole, or potentially an intermediate-mass black hole (IMBH)—that 
accretes material from a companion star. 
These systems can contribute significantly to the X-ray luminosity of galaxies. 
The X-ray point-source emission from relatively young ($\sim$100 Myr)
high-mass X-ray binaries (HMXBs) and older ($\sim$1 Gyr) low-mass X-ray binaries (LMXBs) 
correlates well with galaxy SFR and M$_*$, respectively (\citealt{2003MNRAS.339..793G, 
2004ApJ...602..231C, 2004MNRAS.349..146G}).

We apply the equation (1) to estimate the 
expected X-ray luminosity from XRBs in dwarf galaxies \citep{2010ApJ...724..559L, 2024ApJ...974...14S}.

\begin{equation}
  \rm L_{2-10keV} (erg\, s^{-1}) = \alpha M_* + \beta SFR
\end{equation}

Here, $\alpha$ is (9.05$\pm$0.37)$\times$10$^{28}$ erg s$^{-1}$ M$_{\odot}$$^{-1}$ and 
$\beta$ is (1.62$\pm$0.22)$\times$10$^{39}$ erg s$^{-1}$ (M$_{\odot}$ yr$^{-1}$)$^{-1}$.
The dispersion of this relation is 0.34 dex.

The expected X-ray luminosity (L$_{\rm 2.0-10keV}^{\rm exp}$) is 2.57 $\times$10$^{37}$ erg s$^{-1}$ 
for AGC 223218,
which is two orders of magnitude lower than the observed L$_{\rm 2.0-10keV}^{\rm obs}$.

We listed the above parameters in Table \,\ref{table:agc}.

\begin{table*}[h!tb]
  \caption{Parameters of AGC 223218 and its NSC} 
  \begin{tabular}{c c c c c c c c c c} 
  \hline\hline
   1  & 2 & 3 & 4 & 5 & 6 & 7& 8& 9 & 10 \\ 
  \hline
    &   &  &  &  &  &   &  & & \\ 
  Source &  Velocity$_{\rm Helio}$ & M$_*$  & Mass Ratio &  Age & [Fe/H] & SFR &L$_{\rm 0.2-2.3keV}^{\rm obs}$ & L$_{\rm 2.0-10keV}^{\rm obs}$ & L$_{\rm 2.0-10keV}^{\rm exp}$  \\
    &   &  &  &  &  &   &  & & \\ 
  &   & $\times$10$^7$ & (M$_{\rm *NSC}$/M$_{\rm *gal}$) &  &  &  & $\times$10$^{39}$ & $\times$10$^{39}$ & $\times$10$^{37}$\\
    &   &  &  &  &  &   &  & & \\ 
  &   km s$^{-1}$ & M$_{\odot}$  &  & Gyr &  & M$_{\odot}$ yr$^{-1}$ &  erg s$^{-1}$  &  erg s$^{-1}$ &  erg s$^{-1}$ \\
  \hline  
  &   &  &  &  &  &   &  & & \\ 
  AGC 223218 &  1206$\pm$5 & 5.2$\pm$1.9 $^{\rm SED}$ &  & 1.73  & 0.26  &  0.013$\pm$0.005 $^{\rm SED}$    &  2.44$\pm$1.19  & 2.55$\pm$1.24 & 2.57 \\
   &   &  &  &  &  &  & & & \\
  &   & 12.0$\pm$0.8 $^{\rm MLCR}$ &   &  &  &   & &  &  \\ 
  \hline
  &   &  &  &  &   & &  &  &\\ 
  NSC     &  983$\pm$18   &  0.49$\pm$0.16 $^{\rm SED}$ &   0.094 $^{\rm SED}$  &   0.724  & 0.12 &      &         &  &  \\
   &   &  &  &  &  &  & & & \\
  &   & 0.87$\pm$0.04 $^{\rm MLCR}$ & 0.072 $^{\rm MLCR}$ &  &  &   &   & & \\ 
\hline
  \end{tabular}
  \label{table:agc} 
\end{table*}

\section{Discussions}

\subsection{NSC formation scenario in AGC 223218}

The NSC in AGC 223218 is composed of a younger stellar population (724 Myr), as indicated by the BC03 models. 
GCs are not expected to have formed by so young ages \citep{2013ApJ...775..134V}, 
ruling out the GC accretion scenario for the formation of the NSC in AGC 223218.
The in situ scenario could explain the young stellar population found in NSC.
The primary mechanism in the in situ scenario is gas inflow and star formation at the galactic center.
The gas accretion in the center region indeed discussed for the formation of the core 
in gaint low surface brightness galaxy (GLSBG) Malin 1 \citep{2024A&A...686A.247J}.
However, AGC 223218 is a dwarf LSBG, it may have a diffuse gas distribution and a shallow gravitational potential, 
making in situ star formation less efficient compared to high surface brightness galaxies.
Furthermore, AGC 223218 does not present bar or spiral structures,
which could not support gas inflow mechanisms that promote in situ star formation.

NSC formation via extreme star formation episodes triggered by merger events has been explored 
in several studies \citep{1994ApJ...437L..47M, 2005ApJ...623..650K, 2012MNRAS.421.1927K}. 
\citet{2025ApJ...978L..44R} suggested that mergers could account for the formation of pseudo-bulges in UDGs, 
which are likely NSCs in these systems. Additionally, \citet{2023A&A...677A.179G} and \citet{2024ApJ...971..181C} 
discussed star formation triggered by mergers in those low-mass galaxies.

\cite{2025ApJ...978L..44R} discussed that the merging of dwarf galaxies can trigger central starbursts, 
leading to the formation of massive compact cores dominated by young stellar populations. 
This scenario suggests that these compact cores are typically younger (bluer) and embedded 
within older (redder) disks with a more diffuse spatial distribution. 
In this study, the NSC is predominantly composed of a young stellar population,
with a slightly bluer color compared to AGC 223218, aligning with the findings of \cite{2025ApJ...978L..44R}.
Furthermore, \cite{2025OJAp....8E..90Z} classified post-merger systems of LSBGs using the sample 
from the Systematically Measuring Ultra-Diffuse Galaxies (SMUDGes, \citealt{2019ApJS..240....1Z}) program,
and noted that these systems are highly asymmetric and may host nuclear star clusters.
AGC 223218 fits the classification criteria outlined by \citet{2025OJAp....8E..90Z}.

The merger scenario is able to explain the formation of the NSC in AGC 223218. 
The NSC in AGC 223218 shows lower metallicity than the host galaxy,
which may indicate that the merger event brings pristine gas into AGC 223218.
The merger could explain the offset between the stellar and gas components, 
as well as the higher NSC mass fraction observed in AGC 223218.

We demonstrate that AGC 223218 is located at the boundary between the Seyfert and star-forming regions in the 
[SII]-BPT diagram (Figure\,\ref{BPT}), whereas in the [NII]-BPT diagram, it falls within the star-forming track of SDSS galaxies. 
This suggests the presence of strong shocks in AGC 223218.  
During a merger, gas from both galaxies undergoes strong compression due to tidal forces and gravitational interactions, 
which can trigger intense star formation and generate shock waves.

The merger scenario is considered to explain the formation of extended disks in GLSBGs and the formation 
of superthin galaxies in IllustrisTNG simulation \citep{2018MNRAS.480L..18Z, 2023MNRAS.523.3991Z, 10.1088/1674-4527/ad5399}.
Superthin galaxies are LSBGs seen edge-on \citep{2013MNRAS.431..582B, 2022MNRAS.514.5126N}.
However, the spatial and mass resolution is not sufficient to resolve the NSC in the simulation.

\subsection{Asymmetry of AGC 223218}

Both the asymmetric structure of the galaxy and the clumpy structure around the NSC 
could be caused by a merger event of AGC 223218.

Moreover, \cite{2023Natur.623..296W} 
discussed that ultra-compact dwarf galaxies (UCDs) in the Virgo Cluster form through tidal stripping, 
with their early evolutionary phase characterized as nucleated ultra-diffuse galaxies, 
exhibiting ultra-diffuse tidal features around UCDs. 
In high-density environments, tidal stripping can significantly reshape the morphology of loosely bound material. 

AGC 223218 is in the Local Supercluster \citep{2009A&A...506..677H}.
There are some bright galaxies (e.g. NGC 4251 and NGC 4274) at a few hundred kpc projected distance away from AGC 223218
at a similar redshift. 
It is possible that the AGC 223218 is weakly affected by the Local Supercluster environment. 
The tidal stripping can explain the asymmetrical morphology in the outer region of AGC 223218.
In this scenario, the velocity offset between the NSC and gas components may exist.

\subsection{An IMBH in NSC of AGC 223218}

Due to low SFR and diffuse structure of LSBGs, 
their expected X-ray luminosity is generally lower.
In Section 3.5, the expected X-ray luminosity from X-ray binary contribution 
is significantly lower than the observed X-ray luminosity in AGC 223218,
indicating the presence of a black hole in the NSC. 
Furthermore, AGC 223218 is located at the boundary between 
the Seyfert and star-forming regions in the [SII]-BPT diagram (Figure\,\ref{BPT}).
The x-axis values of both the NSC and AGC 223218 in the [SII]-BPT diagram 
follow the main track of the Seyfert region in the diagram, 
suggesting the possible presence of an AGN.

We estimate the accretion rate of the black hole in AGC 223218
required to account for the X-ray luminosity observed by eROSITA. 
The black hole mass is inferred using the empirical relation between 
host galaxy stellar mass and black hole mass (logM$\rm_{BH}$) for dwarf galaxies in \cite{2015ApJ...813...82R}:
\begin{equation}
  \rm logM_{BH} = 7.45 + 1.05 \times log(M_*/10^{11})
\end{equation}
Using stellar masses derived from SED fitting and MLCR methods, the estimated M$\rm_{BH}$ is 
 1.0 $\times$ 10$^4$ and 2.4 $\times$ 10$^4$ M$\odot$, respectively.
The Eddington luminosity is calculated using the relation in  \cite{2025arXiv250322791K}:
\begin{equation}
 \rm L_{Edd} = 1.26 \times 10^{44} \times (M\rm _{BH}/10^6) (erg\,\,s^{-1})
\end{equation} 
To match the observed X-ray luminosity from eROSITA, the corresponding accretion rates 
are 0.001 and 0.002 for the SED- and MLCR-based stellar masses, respectively.

\cite{2013ARA&A..51..511K} suggested that intermediate-mass black holes (IMBHs) may reside in 
the centers of globular clusters (GCs) and in the NSCs of bulgeless, late-type, and spheroidal galaxies. 
For instance, NGC 4395, a nearby LSBG and the closest known Seyfert 1 galaxy 
hosts a NSC at its center \citep{2013ARA&A..51..511K}.
The central M$\rm_{BH}$ of NGC 4359 is estimated to be in the range of $\sim$ 10$^3$ to 10$^5$ M$_{\odot}$
\citep{2003ApJ...588L..13F, 2005ApJ...632..799P, 2015ApJ...809..101D, 2021ApJ...921...98C, 2023ApJ...948L..23W, 2024MNRAS.530.3578G}.
\cite{2020ApJ...898..106H} reported X-ray nuclei indicative of low-level accretion from massive 
black holes in four nearby LSBGs. 
Consequently, the presence of an IMBH in the NSC of AGC 223218 is not surprising.
However, the X-ray data reveals only a weak detection towards AGC 223218 
due to short exposure and relatively large PSF of eROSITA image. 
The measurement of its X-ray luminosity requires further confirmation.

The morphology of AGC 223218 differs from that of NGC 4395. While NGC 4395 exhibits spiral patterns and 
multiple star-formation regions throughout its disk, AGC 223218 displays only a few clumpy regions around the NSC. 
The fraction of AGN in LSBGs is lower than in their high-surface-brightness counterparts
\citep{2011ApJ...728...74G}.
These two galaxies represent distinct conditions for studying the formation of NSCs and black holes in LSBGs. 
Could dwarf mergers play a significant role in the formation of NSCs and/or black holes in LSBGs? 
Future, high-quality observations, 
including higher resolution HI observations, optical spectra, and X-ray data, 
of AGC 223218 are essential to confirm the findings presented in this study.

\section*{Acknowledgments}
We thank the anonymous referee for a number of very constructive comments.

J.W. acknowledges National Key R\&D Program of China
(grant No. 2023YFA1607904) and the National Natural
Science Foundation of China (NSFC) grants Nos. 12033004,
12333002, and 12221003 and the science research grants
from the China Manned Space Project with Nos. CMS-CSST-2025-A10 and CMS-CSST-2025-A07. 
T.C. acknowledges the China Postdoctoral Science Foundation (grant No. 2023M742929),
and the NSFC grants 12173045 and 12073051.
C.C. is supported by the NSFC, Nos. 11803044, 11933003,
and 12173045 and acknowledges the science research grants
from the China Manned Space Project with No. CMS-CSST-2021-A05. 
G.G. acknowledges the ANID BASAL projects ACE210002 and FB210003.
H.W. acknowledges the NSFC grant Nos. 11733006 and 12090041. 
This work is sponsored (in part) by
the Chinese Academy of Sciences (CAS), through a grant to the
CAS South America Center for Astronomy (CASSACA), NSFC grants Nos. 12090040 and
12090041, and the Strategic Priority Research Program of the
Chinese Academy of Sciences, grant No. XDB0550100.

\bibliographystyle{apj}
\bibstyle{thesisstyle}
\bibliography{main.bib}
\end{document}